# DOUBLE MODELLING OF THE DYNAMIC OF ACTIVITIES IN RURAL MUNICIPALITIES

SÔNIA TERNES[1]
FLORIANA GARGIULO[2]
SYLVIE HUET[3]
GUILLAUME DEFFUANT[4]

**ABSTRACT**: Land use choices and activity prevalence in a selected territory are determined by individual preferences constrained by the characteristic of the analysed zone: population density, soil properties, urbanization level and other similar factors can drive individuals to make different kind of decisions about their occupations. Different approaches can be used to describe land use change, occupation prevalence and their reciprocal inter-relation. In this paper we describe two different kinds of approaches: an agent based model, centred on individual choices and an aggregated model describing the evolution of activity prevalence in terms of coupled differential equation. We use and we compare the two models to analyse the effect of territorial constraints, like the lack of employment in determined sectors, on the possible activity prevalence scenarios.

**KEY-WORDS:** agent based model, population dynamics, rural municipalities, multi-functionality, policy impacts.

## DUPLA MODELAGEM DA DINÂMICA DE ATIVIDADES EM MUNICÍPIOS RURAIS

**RESUMO:** Diferentes formas de uso do solo bem como a prevalência de determinadas atividades de trabalho em territórios são determinadas pela escolhas dos indivíduos restritas às características da zona analisada: densidade populacional, propriedades do solo, nível de urbanização e outros fatores similares podem guiar os indivíduos a tomar diversas decisões sobre a forma de ocupação da terra. Várias abordagens podem ser usadas para descrever mudanças em uso do solo, prevalência de atividades e suas inter-relações. Neste artigo apresentamos duas abordagens distintas: um modelo baseado em agentes, centrado na escolha dos indivíduos, e um modelo matemático analítico que descreve a evolução da prevalência de atividades por meio de um sistema de equações diferenciais. Os dois modelos são testados e comparados para avaliar o efeito de restrições territorial, como a falta de empregos em determinados setores, supondo diferentes cenários de prevalência de atividades.

**PALAVRAS-CHAVE**: modelo baseado em agentes, dinâmica de populações, municípios rurais, multi-funcionalidade, impacto de políticas.

## 1. INTRODUCTION

PRIMA – Prototypical Policy Impacts on Multifunctional Activities in Rural Municipalities – is an European project (FP7-ENV-2007) which aims to develop a framework to enable the impact assessment for European land use policies and land management practices at national and regional levels. One of its goals is to develop a model at municipality level for analysing land use and land cover changes (LUCC).

[1] PhD Electrical Engineering, Embrapa Agricultural Informatics, Post-doc researcher at LISC/Cemagref, France, E-mail: sonia.ternes@cemagref.fr
[2] PhD Physics, Post-doc researcher at LISC/Cemagref, France, E-mail: floriana.gargiulo@cemagref.fr
[3] MsC and Engineer in Computing Engineering, LISC/Cemagref, France, E-mail: sylvie.huet@cemagref.fr
[4] Professor Cognitive Science, LISC/Cemagref, France, E-mail: guillaume.deffuant@cemagref.fr

Several different approaches have been developed to simulate LUCC (Gotts et al., 2003). Each one of these approaches is more or less suitable according to the specific problem, the level of precision required and the considered scale.

Agent based models (ABM) are focused on human actions realized by a set of agents. Agents are autonomous, share an environment through agent communication and interaction and make decisions (for instance like rational optimizers) that connect behaviours to the environment. The dynamics of the whole system is given by the aggregation of all individual behaviours .

Another approach is "microsimulation", which defines individual economic and social trajectories through a set of events which occur with given probabilities (generally neglecting interactions between individuals). Creating a spatial microsimulation model by crossing geographical information with such individual trajectories may provide information about the impact of policies on everyday life and its consequences at the global level (Ballas et al., 2005; Holm et al., 2004).

However such agent based or micro-simulation models can grow rapidly in complexity and become difficult to understand themselves. It is therefore a good methodology to check their behaviour, as much as possible, with approximate simpler models (Deffuant 2004). This paper presents such a work realising in the PRIMA project. It considers the individual dynamics of changing activity patterns in rural areas, aiming to study the relation between demography, activities and the impact of public policies in land use change. In this paper, we study the model without demographic change, to be able to understand better the impact of demography later on. Section 2 describes the microsimulation model while section 3 describes the analytical mathematical model mimicking the microsimulation. Section 4 presents the results of simulations considering the two models, and section 5 concludes and discusses future works.

## 2. THE MICROSIMULATION MODEL

The model represents the population of a municipality or a village where a constant set of individuals change of activities according to some probabilities. The activities can for instance be farming, general services and tourism tasks. We suppose that some of these activities can be limited. For instance, the number of farmers is limited by the land surface available. The individuals can have one or more activities at the same time. Each possible combination of activities represents a specific "activity pattern" which describes an individual occupation.

The transition between activity patterns is given by a probability matrix. At each time step the generalized probability $a_{ij}$ for an individual to change from pattern $i$ to pattern $j$ is given by:

$$a_{ij} = p_{ij} . Av_j$$

where:
- $Av_j$ is the availability for the pattern $j$; it is 1 if the activities composing the pattern are available, 0 if at least one of these activities is not available.
- $p_{ij}$ describes how the transition between the two patterns is probable regarding the individual point of view: $0 \leq p_{ij} \leq 1$. If the value is zero, the transition is not allowed.

The transition occurs only if there is an availability for the pattern, namely if all the activities composing the pattern are available. A Monte Carlo procedure is used to choose the pattern $j$ from the possible values of $a_{ij} > 0$. When an individual has changed of pattern, the availability of the corresponding activities are updated.

Section 4 presents some simulated results of a toy example using the concepts described above considering four activity patterns:

$\Pi_0$ : represents unemployed (no activity);
$\Pi_1$ : represents farming activity;
$\Pi_2$ : represents general services activity;
$\Pi_3$ : represents farming and services at the same time (multi-functionality).

## 3. DENSITY EVOLUTION MODEL

In this section, we describe a reformulation of the microsimulation model as density evolution model. In this framework individuals and their choices are not taken into account. The analysed quantities are directly the densities of occupation of each activity pattern:

$$\overline{\rho} = (\rho_0 \ldots \rho_N) \quad (1)$$

The evolution of the relative occupancy of the patterns is given in terms of coupled differential equations where the essential parameters are derived from the transition matrix between patterns, $p_{ij}$, described in the previous session.

If the effect of availability were neglected, the density of occupants for the *i*-th pattern at time *t*+1 will be given by the balance of the fluxes inside and outside the pattern, namely the proportion of persons that reach the pattern *i* at time *t+1* minus the proportion of persons that leaves from pattern *i* to a new activity pattern at time *t+1*. Since $p_{ij}$ represents the probability of the transition between pattern *i* and pattern *j*, we can write the equation for the evolution of occupancy of the *i*-th pattern in the following way:

$$\frac{d\rho_i}{dt} = \sum_{j=1}^{N} p_{ji} \rho_j - \sum_{j=1}^{N} p_{ij} \rho_i \quad (2)$$

If we consider the effect of availability, we have to set a threshold on the maximum occupancy of each pattern, according to the availability of each activity, which modify the transition probabilities. In the procedure for the introduction of the constraints we must also take into account the fact that the patterns with multiple activities contribute to the saturation of all the activities relative to the pattern: namely, a person that is working in farming and in tourism decreases the accessibility to jobs both in farming and tourism. For sake of simplicity, we specify the simple case of a village with only two possible activities (farming and services). The transition to each of the four patterns should consider the limited availability of the two activities: $Av_1$ and $Av_2$. For realizing this constraint we substitute the transition probabilities $p_{ij}$ with generalized transition probabilities $a_{ij}$ that contain the information about the limited availabilities for the activities. The patterns that contribute to the saturation of activity 1 (farming) are $\Pi_1$ and $\Pi_3$, while the pattern that contribute to the saturation of activity 2 (services) are $\Pi_2$ and $\Pi_3$.

Since the transition to unemployment does not include any availability constraint, i.e. it's always possible to loose an activity, we can set:

$$a_{i0} = p_{i0} \quad (3)$$

For introducing the constraints in the other cases with a continuous function, we use the hyperbolic tangent. The transition to pattern $\Pi_1$, for example, should be modelled using the density of occupation of pattern $\Pi_1$ and $\Pi_3$. The same reasoning is valid for the transition to $\Pi_2$. In such a way we can write:

$$a_{i1,2} = \frac{1}{2} p_{i1,2} \left[ \tanh\left(1 - \frac{\rho_{1,2}(t) + \rho_3(t)}{Av_{1,2}}\right) + 1 \right] \quad (4)$$

The transition to multi-functionality $\Pi_3$, from one of the two activities present in $\Pi_3$, should contain a check on the availability of the other activity:

$$a_{13} = \frac{1}{2} p_{13} \left[ \tanh\left(1 - \frac{\rho_2(t)+\rho_3(t)}{Av_2}\right) + 1 \right] \quad (5)$$

$$a_{23} = \frac{1}{2} p_{23} \left[ \tanh\left(1 - \frac{\rho_1(t)+\rho_3(t)}{Av_1}\right) + 1 \right] \quad (6)$$

The transition from (to) unemployment (pattern $\Pi_0$) to (from) multi-functionality (pattern $\Pi_3$) is zero, since we consider that at each step, only one activity can be reached or lost. Finally, the generalized probability matrix must be normalized as:

$$\tilde{a}_{ij} = \frac{a_{ij}}{\sum_{j=0}^{N} a_{ij}} \quad (7)$$

With this generalized probabilities we can re-write equation (2) in the form:

$$\frac{d\rho_i}{dt} = \sum_{j=1}^{N} \tilde{a}_{ji}(\bar{\rho})\rho_j - \sum_{j=1}^{N} \tilde{a}_{ij}(\bar{\rho})\rho_i \quad (8)$$

## 4. RESULTS

The microsimulation model was simulated from a computational prototype developed in Java programming language. The analytical equations were programmed using the Mathematica software. Figure 1 shows the results of the dynamics of changing patterns, which were obtained under hypothesis of a small village having a constant population of 400 inhabitants and a matrix of transition probabilities $p_{ij}$ given by:

$p_{ij}$ = (0.20 0.10 0.70 0.00,
0.02 0.70 0.00 0.08,
0.09 0.00 0.90 0.01,
0.00 0.28 0.15 0.75).

At each time step, a subset of individuals of the population is selected and they decide to change or not their activity pattern.

The figures show a comparison between the microsimulation and the density model considering an infinite and a limited availability ($Av_1$ = 350 and $Av_2$ = 100). In both cases the results for the microsimulation is represented by continued lines and the results for the analytical models is represented by dotted lines.

For the case of infinite availability (figure 1a) the dynamics of changing activity patterns is determined only by the values of the probability matrix. We can observe that the pattern 2 (services) presents high occupancy during the whole simulation, since the probability of remaining in this patterns is high (0.90), as well the probability of transition from unemployed to services (0.70). The level of unemployment remains at about 10% as well the farming and the multi-functionality patterns.

When the availability of activities decrease (figure 1b), we can observe that the pattern occupancy between farm and services is totally different. This is due to the fact that the activity of service can be only chosen by 25% (proportion of $Av_2$) of the population. Thus the service pattern and the multi-functionality one ("farm + services") are hardly reachable. Then, the rest of the population can choose between the farm pattern, which is widely available, and the unemployment. As the probability to choose the farm pattern is higher than the one to choose unemployment, the majority of the population become a farmer. In both cases showed in the figures 1a and 1b the results obtained from the microsimulation and the analytical model are the same.

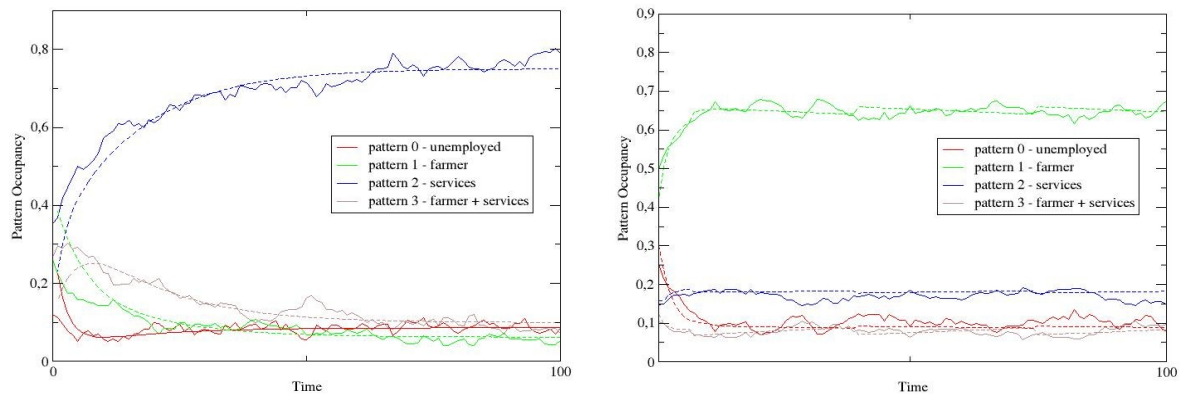

Figure 1: results for the ABM (continuous lines) and the analytical model (dotted lines), considering: (a) infinite availabilities at the left; (b) limited availabilities at right: $Av_1 = 350$ and $Av_2 = 100$.

## 5. CONCLUSIONS

In this paper we presented a microsimulation model of a rural municipality. The model aims to study the individual dynamic of changing activity. This type of model is particularly difficult to study since it has no analytical solution. To better understand its behaviour, we propose and present an density model which approximates this microsimulation model. We show how the aggregated model fits the microsimulation one on two simple examples. The comparison shows that the two models are equivalent for the tested parameters. Then, the aggregated model should be a good tool to study the microlevel model since it gives a reference result. This kind of analysis is a necessary step before adapting the model to more complex situations, for example to analyse the impact of local or global policies on LUCC, the activity prevalence in rural areas and the impact of demographic processes.